\documentclass[twocolumn,superscriptaddress,showkeys]{revtex4-1}
\usepackage{amsmath}
\usepackage{hyperref}
\usepackage{graphicx}
\usepackage{epstopdf}
\usepackage{caption}
\usepackage{subcaption}
\usepackage{float}

\begin{document}
	
	\title{Quantum fluctuations stabilize an inverted pendulum}
	\author{Rohit Chawla} 
	\email{rohit.chawla93@gmail.com}
	\affiliation{School of Mechanical and Materials Engineering, University College Dublin, Belfield, Dublin 4, Ireland}
	\author{Soumyabrata Paul}
	\email{soumyabrata.paul93@gmail.com}
	\affiliation{Department of Physics, Indian Institute of Technology Madras, Chennai 600036, India}
	\author{Jayanta K. Bhattacharjee}
	\email{jayanta.bhattacharjee@gmail.com}
	\affiliation{Department of Theoretical Physics, Indian Association for the Cultivation of Science, Jadavpur, Kolkata 700032, India}
	\date{\today}	
	
	\begin{abstract}
	We explore analytically the quantum dynamics of a point mass pendulum using the Heisenberg equation of motion. Choosing as variables the mean position of the pendulum, a suitably defined generalised variance and a generalised skewness, we set up a dynamical system which reproduces the correct limits of simple harmonic oscillator like and free rotor like behaviour. We then find the unexpected result that the quantum pendulum released from and near the inverted position executes oscillatory motion around the classically unstable position provided the initial wave packet has a variance much greater than the variance of the well known coherent state of the simple harmonic oscillator. The behaviour of the dynamical system for the quantum pendulum is a higher dimensional analogue of the behaviour of the Kapitza pendulum where the point of support is vibrated vertically with a frequency higher than the critical value needed to stabilize the inverted position. A somewhat similar phenomenon has recently been observed in the non equilibrium dynamics of a spin - $1$ Bose-Einstein Condensate. 
	\end{abstract}
	
	
	\maketitle

	\section{Introduction} \label{1}

	The different aspects of quantum dynamics of a point mass pendulum has rarely been the subject of investigation. Cook and Zaidins \cite{1} estimated the time and fall of a pendulum due to the quantum fluctuations. Doncheski and Robinett \cite{2} discussed in detail the limiting cases of a simple harmonic oscillator and the free rotor and focussed on the issues of the wave packet revival. Leibsher and Schmidt \cite{3} carried out a detailed numerical investigation. However, recently a completely new aspect was noticed by Gerving et al. \cite{4} who focussed on the dynamics with initial conditions set near the unstable fixed point of the classical pendulum.
	
	\vspace{.1in}	
	
	Gerving et al \cite{4} studied the non equilibrium dynamics of a spin-$1$ Bose-Einstein Condensate initialized in an unstable state which is analogous in the mean field limit to the exactly inverted pendulum. They have measured the evolution of this state along a separatrix caused by quantum fluctuations. Subsequently in a thesis submitted to the physics department of Georgia Institute of Technology, Gerving \cite{5} presents in chapter 7 a semi classical calculation of the motion of a wave packet which is initially centred around the unstable equilibrium point of the pendulum. Surprisingly, the mean angular displacement of the quantum pendulum shows oscillation around the unstable equilibrium point. The variance has oscillatory behaviour as well and there is a marked skewness which also oscillates in time. Clearly the quantum fluctuations are \textit{stabilizing} the unstable fixed point of the classical pendulum. This is reminiscent of the Kapitza pendulum \cite{6} where a high frequency vertical vibration of the point of support stabilizes the unstable position. Gerving's computation is actually based on an approach using an ensemble of classical orbits as advocated by Ballentine et al \cite{7} and explored by several authors \cite{8,9,10,11}. In this work, we use the average angular displacement, a suitably defined generalised variance and a generalised skewness as the dynamical variables to set up a dynamical system approach which clearly establishes the \textit{stabilization} of the classical unstable point by angular fluctuations. Our approach is a variation on the theme of dynamics of moments introduced recently by Brizuela \cite{12,13}. 
	
	\vspace{.1in}	
	
	The pendulum is described by the angular variable $\theta$ and the conjugate momentum $p_\theta$. We start with Heisenberg equations of motion for any operator $O$ and take an expectation value in any quantum state $\psi(\theta,t)$ to write\cite{9},\cite{10}.
	
	\begin{equation} \label{eq:1}
		\imath\hbar \frac{d}{dt} \langle \hat{O} \rangle = \langle [\hat{O},H]\rangle + \imath\hbar \Big\langle \frac{\partial \hat{O}}{\partial t} \Big\rangle
	\end{equation}
	
	where,
	
	\begin{equation} \label{eq:2}
		H = \frac{p_\theta^2}{2m} + m\omega^2 l^2 (1 - \cos\theta )
	\end{equation}

	and
	
	\begin{equation} \label{eq:3}
		\langle \hat{O} \rangle = \int_{-\pi}^{\pi} \psi^*(\theta,t) \hat{O} \psi(\theta,t) d\theta
	\end{equation}
	
	We will henceforth set the mass $m$ and the length of the pendulum to unity and will restore them at the end of the calculation. We will denote the average angular displacement of the pendulum for a quantum state to be $\phi$ (i.e $\phi = \langle \theta \rangle = \int_{-\pi}^{\pi} \psi^*(\theta,t) \hat{\theta} \psi(\theta,t) d\theta)$  and two applications of Eq. (\ref{eq:1}) leads to
	
	\begin{equation} \label{eq:4}
	\frac{d^2\phi}{dt^2} + \omega^2 \langle \sin \theta \rangle = 0
	\end{equation}		
	
	As is obvious the mean position does not follow the classical trajectory. Writing 
	
	\begin{equation} \label{eq:5}
	\begin{split}	
		\langle \sin \theta \rangle &= \langle \sin (\theta - \phi + \phi) \rangle \\ &=
		\langle \cos (\theta - \phi) \rangle \sin \phi + \langle \sin (\theta - \phi) \rangle \cos \phi
	\end{split}
	\end{equation}
	 
	 we find,
	 
	 \begin{equation} \label{eq:6}
	 \begin{split}
	 	\frac{d^2 \phi}{dt^2} = &- \omega^2 \sin \phi + \omega^2 \langle 1 - \cos (\theta - \phi) \rangle \sin \phi \\ 
	 	&- \omega^2 \langle \sin (\theta - \phi) \rangle \cos \phi
	 \end{split}
	 \end{equation}
	 
	The last two terms on the left hand side represent the effect of quantum fluctuations on the dynamics of the mean angular displacement. We define a generalised variance
	
	\begin{equation} \label{eq:7}
		V = \langle 1 - \cos (\theta - \phi) \rangle
	\end{equation}
	
	and a generalised skewness
	
	\begin{equation} \label{eq:8}
		S = \langle \sin (\theta - \phi) \rangle	
	\end{equation}
	
	For small fluctuations about the mean, $V \simeq  \frac{\langle(\theta - \phi)^2\rangle}{2}$ and $S \simeq  - \frac{1}{6}\langle (\theta - \phi)^3\rangle$, which are the usual definitions of variance and skewness apart from the factors of $1/2$ and $1/6$. For discussion of quantum dynamics in non periodic situations using the usual variance $(\langle \theta^2 \rangle - \langle \theta \rangle^2)$ and subsequent Gaussian approximations, one should consult Ref. \cite{8,9,10,11}. A slightly different point of view is to be found in \citep{16}. We will generalise the technique of \cite{16} to the present situation where small angle approximations are not permitted. In Section \ref{2}, we write down a closed dynamical system for $\phi$, $V$ and $S$ and study it's fixed points and dynamics. The numerical results are shown in Section \ref{3} and we conclude with a discussion in Section \ref{4}.	
	
	\section{The Dynamical System} \label{2}
	
	We need to find the dynamics of $V$ and $S$ without making any small angle approximation. This calls for repeated applications of Eq. (\ref{eq:1}). The issue to be settled is that of the appearance of higher moments. We have used a factorizing scheme, where any correlations of the form $\langle[1-\cos(\theta-\phi)]^2\rangle$, $\langle \sin(\theta-\phi)[1-\cos(\theta-\phi)]\rangle$, $\langle p_\theta^2(1-\cos(\theta-\phi))\rangle$ has been replaced by $V^2, SV, p_\theta^2 V$ etc. and care has been taken to ensure that the simple harmonic oscillator limit is correct. Since the Hamiltonian is a constant of motion, we use, 
	
	\begin{align} \label{eq:9}
		e &= \frac{\langle p_\theta^2 \rangle}{2} + \omega^2 \langle 1-\cos\theta \rangle \\
		  &= \frac{\langle p_\theta^2 \rangle}{2} + \omega^2 (1 - (1-V)\cos\phi + S \sin\phi) \nonumber	
	\end{align}	 
	
	as a parameter of the problem. Long but straightforward algebra now leads to
	
	\begin{equation} \label{eq:10}
	\begin{split}
		\frac{d^2V}{dt^2} = (&2e - 2\omega^2 - \frac{\hbar^2}{4})(1 - V) \\ 
		&+ \omega^2(2-6V+3V^2)\cos\phi \\ 
		&- 2\omega^2S(1-V)\sin\phi - 3\omega^2 S \sin\phi \\
		&+\omega^2 S^2 \cos\phi -2\dot{\phi}\dot{S} - (1 - V)\dot{\phi}^2
	\end{split}
	\end{equation}
	
	and 
	
	\begin{equation} \label{eq:11}
	\begin{split}
		\frac{d^2S}{dt^2} = -(&2e - 2\omega^2 - \frac{\hbar^2}{4})S \\
		&- 5\omega^2 S(1 - V) \cos\phi \\
		&+ 2\omega^2 S^2\sin\phi + 2\dot{\phi}\dot{V} + S\dot{\phi}^2
	\end{split}
	\end{equation}			
	
	Our dynamical system comprises of Eqs. (\ref{eq:6}), (\ref{eq:10}), and (\ref{eq:11}). We should point out that in the limit of very low energies i.e. $e/\omega^2 << 1$, the system behaves like a simple pendulum and for $\omega \rightarrow 0$, it is like a free rotor.
	
	We now look at the fixed points of our dynamical system. There are three of them.
	
	$A\big) \phi^* = S^* = 0, V^* = V_o$.
		
	The value of $V_o$ is found to be,
	
	\begin{equation} \label{eq:12}
	\begin{split}
		6\omega^2V_o = 2e &+ 4\omega^2 - \frac{\hbar^2}{4} \\
					   \pm \Bigg[ &\Big(2e + 4\omega^2 - \frac{\hbar^2}{4}\Big)^2 \\
					   &-12\omega^2\Big(2e - \frac{\hbar^2}{4}\Big) \Bigg]^{1/2}
	\end{split}
	\end{equation}
	
	$B\big) \phi^* = \pi, S^* = 0, V^* = \overline{V_o}$.	
	
	The value of $\overline{V_o}$ is found to be,
	
	\begin{equation} \label{eq:13}
	\begin{split}
		6\omega^2 \overline{V_o} = 8\omega^2 &- 2e + \frac{\hbar^2}{4} \\
						           \mp \Bigg[&\Big(2e - 8\omega^2 - \frac{\hbar^2}{4}\Big)^2 \\
				                             &+ 12\omega^2\Big(2e - 4\omega^2 - \frac{\hbar^2}{4}\Big)\Bigg]^{1/2}
	\end{split}
	\end{equation}
	
	$C\big)$ The third fixed point is found to be,
	
	\begin{equation} \label{eq:14}
		\phi^* \simeq \frac{\pi}{2}, \overline{V_o}^* \simeq 1, S^* < 0
	\end{equation}		

	We see from Eqs. (\ref{eq:12}) and (\ref{eq:13}) that there are two possible values of $V_o$ and $\overline{V_o}$. The choice of the relevant value is made by analysing the stability properties of the fixed points. Linearising about the fixed point $A$, we get,
	
	\begin{subequations}
	\begin{align}
	\frac{d^2\delta \phi}{dt^2} &= -\omega^2(1 - V_o)\delta\phi - \omega^2\delta S \label{eq:15a} \\
	\frac{d^2\delta V}{dt^2} &= -\Big(2e + 4\omega^2 - 6\omega^2 V_o - \frac{\hbar^2}{4}\Big)\delta V \label{eq:15b} \\ 
	\frac{d^2\delta S}{dt^2} &= - \Big(2e - 2\omega^2 + 5\omega^2(1 - V_o) - \frac{\hbar^2}{4}\Big)\delta S \label{eq:15c}                         
	\end{align}
	\end{subequations}
	
	For $\delta V$ to execute small oscillations about $V_o$ as a mark of stability, it is essential that we choose the negative sign in Eq. (\ref{eq:12}). An identical argument for the fixed point ($B$) reveals that the positive sign is relevant for Eq. (\ref{eq:13}).
	
	\vspace{.1in}	
	
	We want to check that our dynamical system does reproduce the known quantum dynamics in the limits of $e >> \omega^2$ and $e << \omega^2$. For $e << \omega^2$, we have a simple harmonic oscillator. The fixed point variance $V_o$ has the value $\frac{e}{2\omega^2}$. In terms of $\Delta^2 = \langle \theta^2 \rangle - \langle \theta \rangle^2$, we have the variable $V \simeq \frac{\Delta^2}{2}$. In the $\omega >> e$ limit, the dynamics of $V$ is given by $V = V_o + \delta V$, where $\delta V$ satisfies
	
	\begin{equation} \label{eq:16} 
		\delta\ddot{V} + 4\omega^2\delta V = 0
	\end{equation}
	 
	 and hence in terms of the quantity $\Delta^2$, we have,
	 
	 \begin{equation} \label{eq:17} 
	 	\Delta^2 = \frac{e}{\omega^2} + A\cos2\omega t + B\sin2\omega t
	 \end{equation}
	 
	If we take an initial Gaussian wave packet having width $\Delta_o$, i.e. $\psi(x,t=0)= \frac{1}{\pi^\frac{1}{4}\Delta_o^2}\exp^\frac{-(x-a)^2}{2\Delta_o^2}$, then $\frac{e}{\omega^2} = \frac{\hbar^2}{4\Delta_o^2\omega^2} + \frac{\Delta_o^2}{4}$. With such a packet $\frac{d\Delta^2}{dt}=0$ at $t = 0$ which makes $B = 0$. Hence $A + \frac{e}{\omega^2} = \frac{\Delta_o^2}{2}$ and Eq. (\ref{eq:16}) becomes
	
	\begin{align} 
		\Delta^2 &= \frac{e}{\omega^2} + (\frac{\Delta_o^2}{2} - \frac{e}{\omega^2})\cos2\omega t \nonumber \\
		&= \frac{e}{\omega^2} + (\frac{\Delta_o^2}{4} - \frac{\hbar^2}{4\Delta_o^2\omega^2})\cos2\omega t	\label{eq:18}
	\end{align}
	
\begin{figure}
	\begin{subfigure}{.5\textwidth}
		\includegraphics[width=\linewidth]{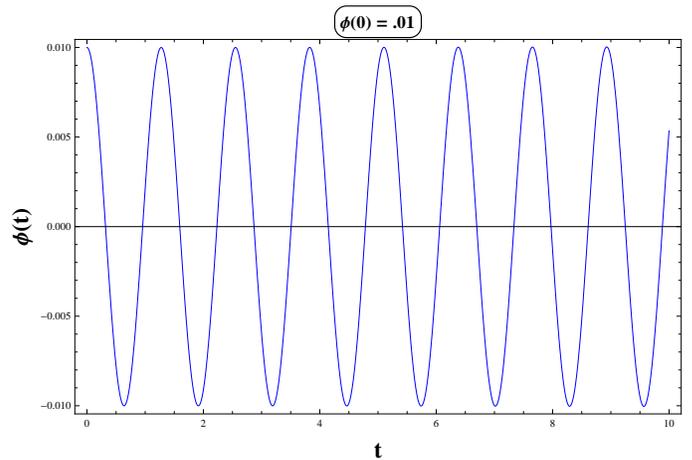}	
		\caption{Plot of $\phi(t)$ with $\phi(0) = .01$ and $\dot{\phi}(0)= 0$}
		\label{Fig 1a}
	\end{subfigure}
	\begin{subfigure}{.5\textwidth}
		\includegraphics[width=\linewidth]{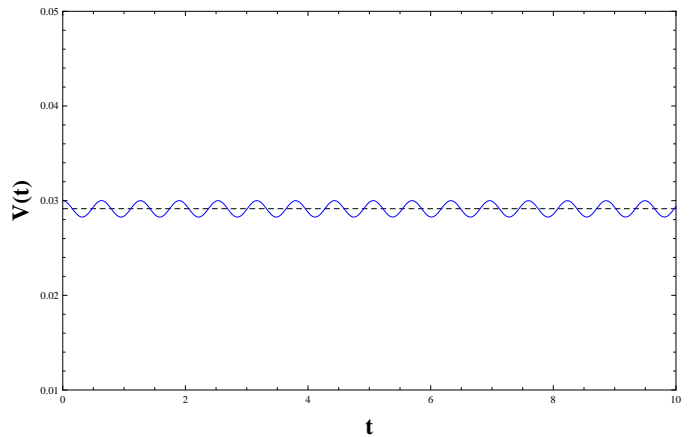}	
		\caption{$V(t)$ with $V(0)=.03$ and $\dot{V}(0)=0$. The dashed line is the fixed point $.02915$ which is a close to $.03$, the value $\frac{e}{2m^2}$ in the harmonic oscillator limit.}
		\label{Fig 1b}
	\end{subfigure}
	\caption{Numerics for $\phi(t)$ and $V(t)$ with $e = 1.5$ and $\omega = 5$}
	\label{Fig 1}
	\end{figure}

	Choosing $\Delta_o^4 = \frac{\hbar^2}{\omega^2}$ (in $m = 1$ units), we find that $\Delta^2$ remains fixed in time with the initial width. This is exactly as it should be since $\Delta_o^2 = \frac{\hbar}{\omega}$ corresponds to the coherent state. Hence the dynamics of the large $\omega$ limit correctly reproduces the essential feature of that limit. This is shown in Fig. \ref{Fig 1}. For the numerics shown, the term $\frac{\hbar^2}{4}$ is a small quantity in comparison with the energy, $e$. Inserting back the dimensional quantities we are comparing $\omega^2$ with $\frac{\hbar^2}{4m^2l^4}$. Choosing it as $1/16$, we proceed with the numerics in the harmonic oscillator limit and all subsequent plots. 

	\vspace{.1in}
	
	We now turn to the free rotor limit, where $\omega \rightarrow 0$. From Eq. (\ref{eq:6}) this leads to, as expected, 
	
	\begin{equation}	\label{eq:19}
		\phi = A_1 t + A_2
	\end{equation}
	
	where $A_1$ and $A_2$ are constants of integration with $A_1$ corresponding to the initial value of $\langle p_\theta^2 \rangle$ which is a conserved quantity. If we analyse the variance $V$, we notice that Eq. (\ref{eq:10}) reduces to $\ddot{V} = 2e^\prime(1 - V)$, with $e$ being redefined as $e^\prime = e - \hbar^2/8$, the solution being,
	
	\begin{equation}	\label{eq:20}
		V = 1 + A\cos(2e^\prime)^{0.5} t + B\sin(2e^\prime)^{0.5} t
	\end{equation}		
	
	Once again for an initially real wave packet, $B = 0$, and if $V = V_o$ at $t = 0$, then,
	
	\begin{equation}	\label{eq:21}
		V = V_o + (1 - V_o)(1 - \cos(2e^\prime)^{0.5} t)	
	\end{equation}		
	
\begin{figure}
	\begin{subfigure}{.5\textwidth}
		\includegraphics[width=\linewidth]{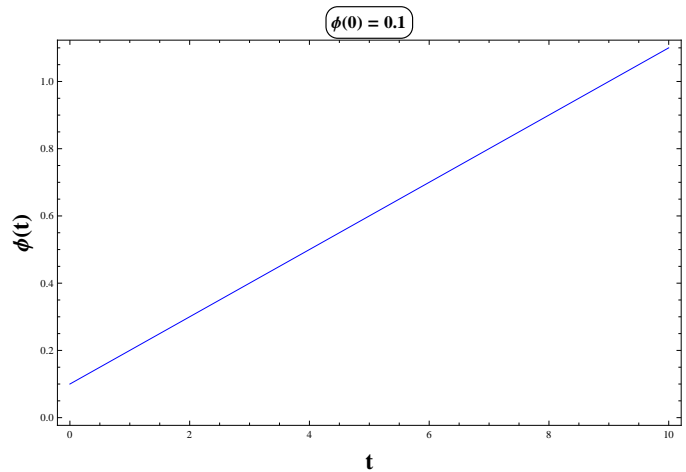}	
		\caption{Plot of $\phi(t)$ with $\phi(0) = .1$ and initial momentum $\dot{\phi}(0) = .1$}
		\label{Fig 2a}
	\end{subfigure}
	\begin{subfigure}{.5\textwidth}
		\includegraphics[width=\linewidth]{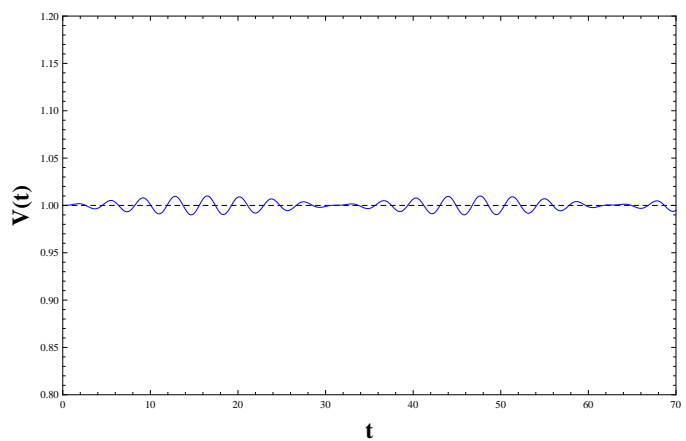}	
		\caption{$V(t)$ with $V(0) = 1$ and $\dot{V}(0)= 0$. $V(t)$ shows small fluctuations about its fixed point value of $1$.}
		\label{Fig 2b}
		\end{subfigure}
	\begin{subfigure}{.5\textwidth}
		\includegraphics[width=\linewidth]{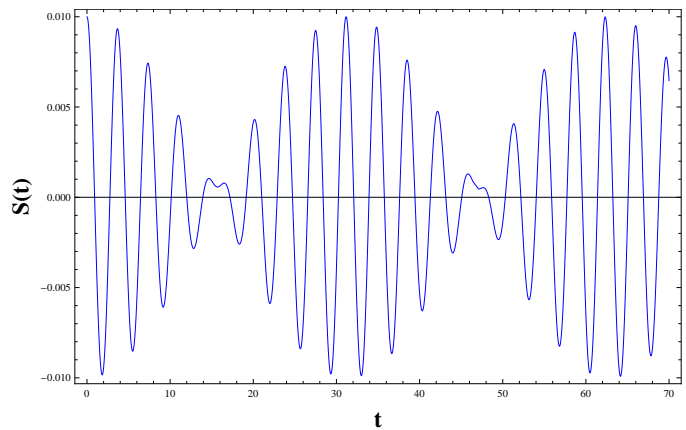}	
		\caption{$S(t)$ with $S(0) = 0.01$ and $\dot{S}(0) = 0$. $S(t)$ oscillates about zero with amplitude $0.01$ }
		\label{Fig 2c}
	\end{subfigure}
	\caption{Numerics for $\phi(t), V(t), S(t)$ with $e = 1.5, \omega = 0$ and initial momentum $\dot{\phi}(0) = .1$}
	\label{Fig 2}
	\end{figure}

	If the initial width $V_o$ is small(large energy), then clearly for times $t << \sqrt{\frac{1}{2e^\prime}}$, $V = V_o + \frac{(1 - V_o)}{2}(2e^\prime)t^2 \simeq V_o + \frac{\hbar^2t^2}{4V_o}$, since $e^\prime \simeq e = \frac{\langle p_\theta^2 \rangle}{2} = \frac{\hbar^2}{4V_o}$ which is the standard free particle limit. Fig. (\ref{Fig 2}) shows the corresponding plots of Eqs. (\ref{eq:6}), (\ref{eq:10}) and (\ref{eq:11}) in the free particle limit. From Fig. (\ref{Fig 2a}), one sees that $\phi(t)$, with an initial momentum, increases linearly with time while $V(t)$ oscillates about $1$ and $S(t)$ shows oscillation about zero. The numerical solution of Eqs. (\ref{eq:6}),(\ref{eq:10}) and (\ref{eq:11}) show the expected behaviour in Fig. \ref{Fig 2}. Confident that the system works in the limiting situations, we now turn to the region of interest where the average energy is of $O(\omega^2)$.    

	\begin{figure}
	\begin{subfigure}{.5\textwidth}
		\includegraphics[width=\linewidth]{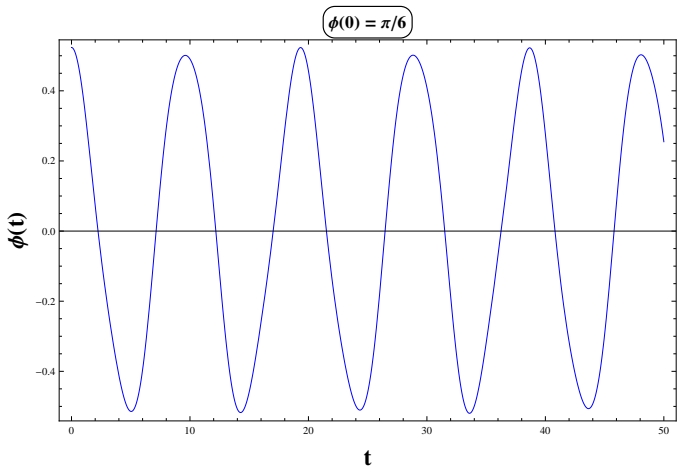}	
		\caption{Plot of $\phi(t)$ with $\phi(0) = \frac{\pi}{6}$ and $\dot{\phi}(0) = 0$}
		\label{Fig 3a}
	\end{subfigure}	
	\begin{subfigure}{.5\textwidth}
		\includegraphics[width=\linewidth]{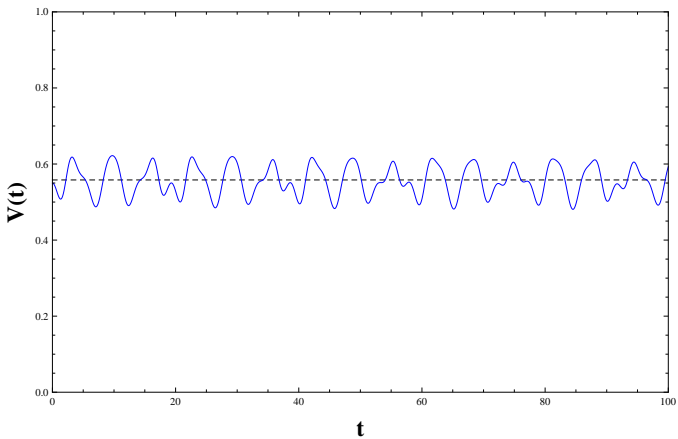}	
		\caption{$V(t)$ with $V(0) = .55$ and $\dot{V}(0) = 0$. The dashed line is the fixed point $V_o = .55813$.}
		\label{Fig 3b}
	\end{subfigure}
	\begin{subfigure}{.5\textwidth}
		\includegraphics[width=\linewidth]{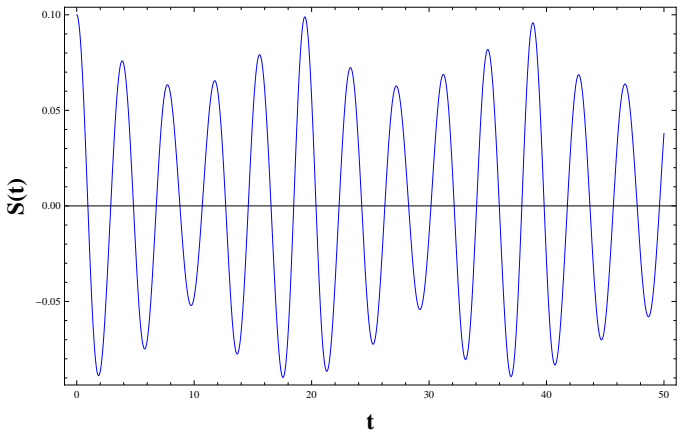}	
		\caption{$S(t)$ with $S(0) = 0.1$ and $\dot{S}(0) = 0$}
		\label{Fig 3c}
	\end{subfigure}
	\caption{Numerics for $\phi(t),V(t)$ and $S(t)$ about the fixed point $A$ with $e = 1.5$ and $\omega = 1$.}
	\label{Fig 3}
	\end{figure}			
	
	\section{Numerical Solution of the Dynamical System} \label{3}
	
	Our three-variable system requires the fixed points for the mean,variance and skewness. For the usual fixed point(i.e. fixed point $A$) where the mean and skewness have value zero, the variance for any given value of $\omega$ has the maximum possible fixed point value of unity. Over the entire range of positions of the fixed point, the dynamics is stable. Our interest in this section is primarily in the fixed point $B$ where the value of the mean is $\pi$(the inverted position of the pendulum), the skewness is zero and the variance is given by Eq. \ref{eq:13}(only the positive sign is relevant because that corresponds to stability). Here for a given $\omega$, the minimum possible fixed point value of variance,V is unity. This is what stabilises the fixed point and creates the unexpected behaviour that the quantum pendulum can execute bounded motion around the classically unstable fixed point provided the variance of the initial wave packet centred near $\phi = \pi$ has a value greater than unity.
	
	\vspace{.1in}	
	
	We first examine the fixed point $A$ with $\omega = 1$ and $e = 3/2$. The results of numerically integrating Eqs. (\ref{eq:6}),(\ref{eq:10}) and (\ref{eq:11}) are shown in Fig. \ref{Fig 3}. We take a mean initial angular displacement of $\pi/6$ radians. The initial variance is given near the fixed point Eq. (\ref{eq:12}), the negative value of $V_o$ being the relevant one here, and a small initial skewness. The derivatives of $\phi$, $V$ and $S$ are specified as zero at $t=0$. The results, as expected show stable oscillations about $\phi = 0$, $V = .55813$(the fixed point value corresponding to $e$ and $\omega$) and $S = 0$.	
				
	\vspace{.1in}				
				
	We now turn to the \textit{unstable} fixed point of the classical pendulum. In our case this is the fixed point $\phi = \pi$, $V = \overline{V_o}$ and $S = 0$ with the positive sign in Eq. (\ref{eq:13}) being the relevant one. The dynamics is stable (i.e. periodic around $\overline{V_o}$) for this choice of sign in Eq. (\ref{eq:10}) as can be found by performing stability analysis with Eqs. (\ref{eq:6}) and (\ref{eq:11}). For $\phi = \pi + \delta\phi$ in Eq. (\ref{eq:6}) and $S = \delta S$ in eq. (\ref{eq:11}), we have,
	
	\begin{align} 
	\frac{d^2\delta \phi}{dt^2} &= \ \omega^2(1 - \overline{V_o})\delta \phi + \omega^2\delta S  \label{eq:22} \\
	\frac{d^2\delta S}{dt^2} &= \ - \Big(2e - 2\omega^2 - \frac{\hbar^2}{4} + 5\omega^2(\overline{V_o} - 1)\Big)\delta S\label{eq:23}								
	\end{align}
	
\begin{figure}
	\begin{subfigure}{.5\textwidth}
		\includegraphics[width=\linewidth]{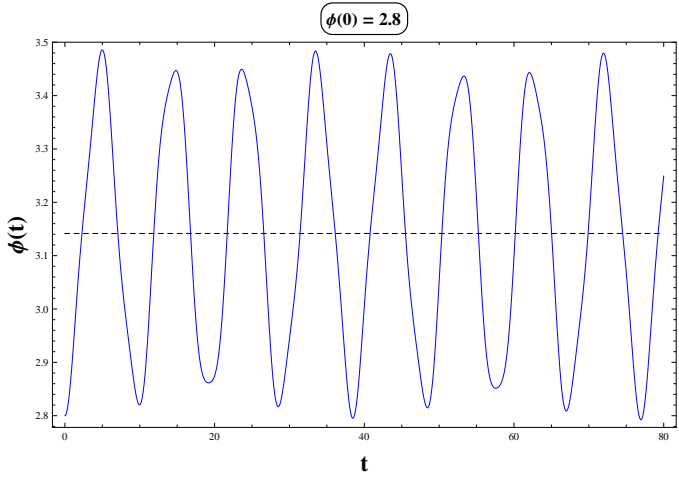}	
		\caption{Plot of $\phi(t)$ with $\phi(0) = 2.8$ and $\dot{\phi}(0) = 0$. The dashed line is the classical unstable fixed point $\pi$.}
		\label{Fig 4a}
	\end{subfigure}	
	\begin{subfigure}{.5\textwidth}
		\includegraphics[width=\linewidth]{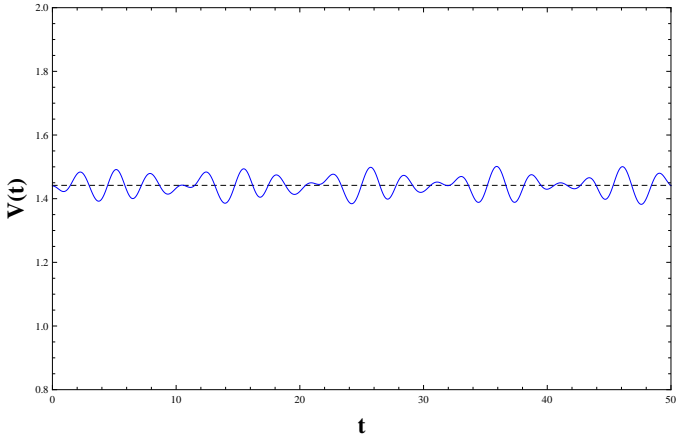}	
		\caption{$V(t)$ with $V(0) = 1.44$ and $\dot{V}(0) = 0$. The dashed line is the fixed point $\overline{V_o} = 1.44187$.}
		\label{Fig 4b}
	\end{subfigure}
	\begin{subfigure}{.5\textwidth}
		\includegraphics[width=\linewidth]{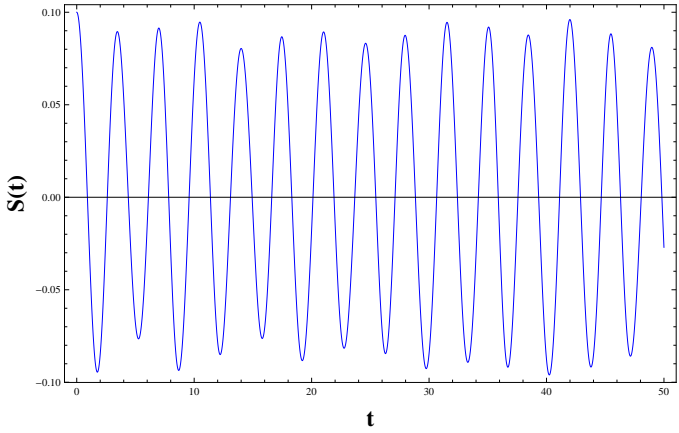}	
		\caption{$S(t)$ with $S(0) = 0.1$ and $\dot{S}(0) = 0$.}
		\label{Fig 4c}
	\end{subfigure}
	\caption{Numerics for $\phi(t),V(t)$ and $S(t)$ about the fixed point $B$ with $e = 1.5$ and $\omega = 1$.}
	\label{Fig 4}
	\end{figure}

\begin{figure}
	\begin{subfigure}{.5\textwidth}
		\includegraphics[width=\linewidth]{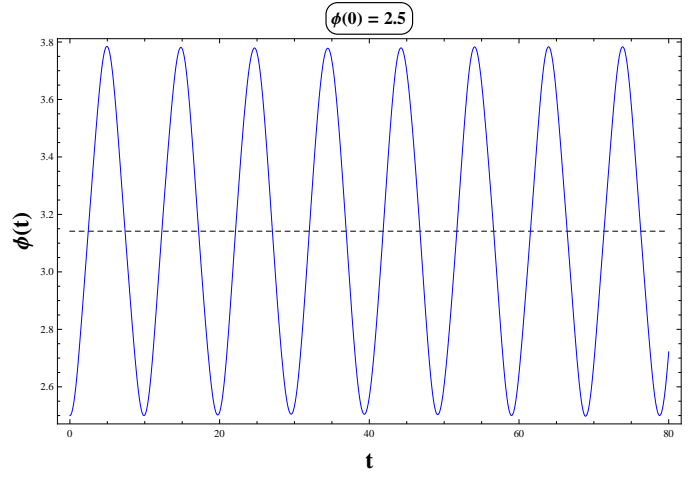}	
		\caption{Plot of $\phi(t)$ with $\phi(0) = 2.5$.}
		\label{Fig 5a}
	\end{subfigure}	
	\begin{subfigure}{.5\textwidth}
		\includegraphics[width=\linewidth]{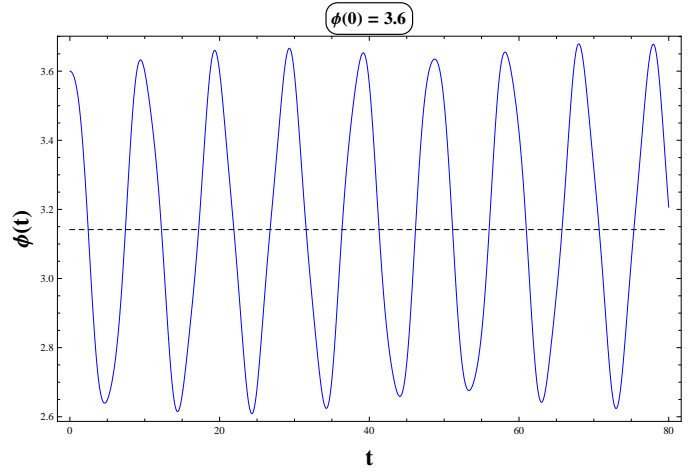}	
		\caption{$\phi(t)$ with $\phi(0) = 3.6$.}
		\label{Fig 5b}
	\end{subfigure}
	\begin{subfigure}{.5\textwidth}
		\includegraphics[width=\linewidth]{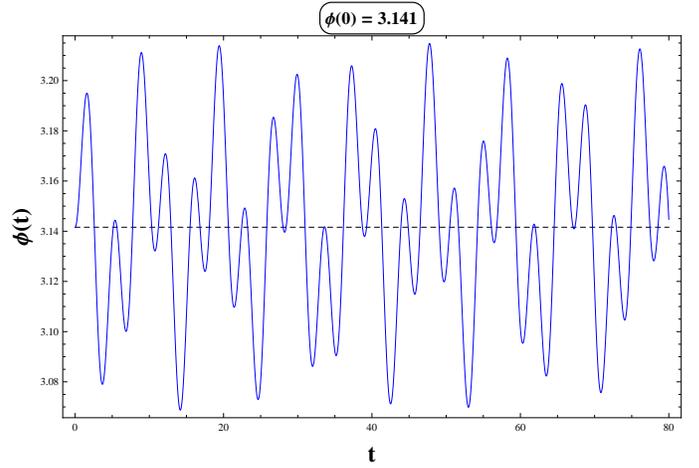}	
		\caption{$\phi(t)$ with $\phi(0) = 3.141$.}
		\label{Fig 5c}
	\end{subfigure}
	\caption{Numerics for $\phi(t)$ for various initial conditions around $\pi$.}
	\label{Fig 5}
	\end{figure}

	The stability matrix has a negative trace and a positive determinant which ensures that the fixed point is a centre and the dynamics about the vertical i.e. upside down position is oscillatory. We again integrate Eqs. (\ref{eq:6}),(\ref{eq:10}) and (\ref{eq:11}) but with initial conditions that should be within the basin of attraction of the fixed point $B$. In Figs. (\ref{Fig 4}) and (\ref{Fig 5}) we show the existence of oscillation about the vertical position. This is the unexpected feature of the quantum pendulum, stemming presumably from the special status \cite{17} of $\phi = \pi$ where the classical system is aperiodic. Fig. (\ref{Fig 4}) describes the nature of $\phi(t), V(t)$ and $S(t)$ with $\phi(0) = 2.8$. Fig. (\ref{Fig 5}) shows the nature of $\phi(t)$ when its released on either sides of $\pi$ and one very close to $\pi$.	

	\vspace{.1in}

	A linear stability around the fixed position $C$ of Eq. (\ref{eq:14}) shows the fixed point to be unstable. Thus the fixed point structure representing the quantum pendulum has an identical structure to that of the Kapitza pendulum so far as the dynamical system analysis goes. We conclude in Section \ref{4} with the discussion of this analogy.		
	
	\section{Conclusion} \label{4}	
		
	In this concluding section, we recall the Kapitza pendulum and point out the similarities and the differences between the two systems. The Kapitza pendulum has its point of support vibrated at a high frequency $\Omega$, so that the equation of motion is given by,
	
	\begin{equation} \label{eq:24}
	\ddot{\theta} + \omega^2(1 + \epsilon\cos{\Omega t})\sin\theta = 0	
	\end{equation}
	
	where $\omega = \sqrt{\frac{g}{l}}$, is the natural frequency of the pendulum. The dynamics can be split into a \textit{slow} (frequency of $O(\omega)$) and a \textit{fast} (frequency of $O(\Omega)$) as $\theta = \theta_s + \theta_f$, with the high frequency oscillation taken to be a small perturbation around the primary solution $\theta_s$. For $\theta_f << \theta_s$, $\ddot{\theta_s} + \omega^2\theta_s = 0$ is the primary dynamics. The fast variation satisfies, to the lowest order, the dynamics 
	
	\begin{equation} \label{eq:25}
	\ddot{\theta_f} + \omega^2 \cos{\theta_s} \ \theta_f = - \ \epsilon \omega^2 \cos{\Omega t} \sin\theta_s ,
	\end{equation}
	
	leading to the approximate solution $\theta_f \simeq \frac{\epsilon \omega^2}{\Omega^2}\sin\theta_s$. Inserting this $\theta_f$ back into Eq. (\ref{eq:25}), leads to the $O(\epsilon^2)$ dynamics of $\theta_s$ as,
	
	\begin{equation} \label{eq:26}
	\ddot{\theta_s} + \omega^2 \sin\theta_s + \frac{\epsilon^2 \omega^4}{2\Omega^2} \sin{2\theta_s}= 0.
	\end{equation}		 		
	
	The effective potential for this dynamics is, 
	
	\begin{equation} \label{eq:27}
	V_{eff} = -\cos\theta_s - \frac{\epsilon^2\omega^4}{4\Omega^2} \cos{2\theta_s}.
	\end{equation}
	
	The fixed point of Eq. (\ref{eq:27}) for $\frac{\epsilon^2\omega^2}{\Omega^2} > 1$ are clearly $\theta_s = 0, \theta_s = \pi$ and $\cos\theta_s = -\frac{\Omega^2}{\omega^2\epsilon^2}$. The stability can be understood from the extrema of $V_{eff}$. The extremum $\theta_s = 0$ is clearly a minimum. The extremum $\theta_s = \pi$ is a minimum for $\frac{\epsilon^2\omega^2}{\Omega^2} > 1$ which corresponds to the stabilization of the inverted position. In this situation, the third fixed point $\theta_s = \cos^{-1}{\Big(-\frac{\Omega^2}{\omega^2 \epsilon^2}\Big)}$ exists and is easily seen to be unstable. 

	\vspace{.1in}
	
	In some ways, our situation is analogous to the Kapitza pendulum because the fixed point structure is similar. However, there are very strong differences. The Kapitza system is a two dimensional non-autonomous system while this quantum pendulum is a six dimensional autonomous dynamical system. The fixed points $A$ and $B$ are correctly two fixed curves in the $e - \omega$ parameter space. While the non autonomous Kapitza system has two distinct time scales as shown by our analysis above, the autonomous quantum oscillator has two separate stable segments ($V^* < 1$ and $V^* > 1$) separated by an unstable region which separates the basins of attraction of the two fixed points.	
	
	\vspace{.1in}	
	
	The above is best understood by analysing fixed points $(A)$ and $(B)$ as a function of $e$ for a given value of $\omega$. We note that for the fixed point $(A)$ where $\phi = 0$, the  fixed point $V_o$ is seen to be (from Eq. (\ref{eq:12}))
	
	\begin{subequations}
	\begin{align}
		V_o &= \frac{e}{2\omega^2} + O\Big(\frac{e^2}{\omega^4}\Big) \ \text{for} \ e << \omega^2 \\ 
		V_o &= 1 - O\Big(\frac{\omega^2}{e}\Big) \ \text{for} \ e >> \omega^2 
	\end{align}	
	\end{subequations}
	
	On the other hand, the fixed point $(B)$ has the form,
	
	\begin{subequations}
	\begin{align}
		\overline{V}_o &= 2 - O\Big(\frac{e}{\omega^2}\Big) \ \text{for} \ e << \omega^2 \\ 
		\overline{V}_o &= 1 + O\Big(\frac{\omega^2}{e}\Big) \ \text{for} \ e >> \omega^2 
	\end{align}	
	\end{subequations}	 
	
\begin{figure}[h]
	\includegraphics[width=\linewidth]{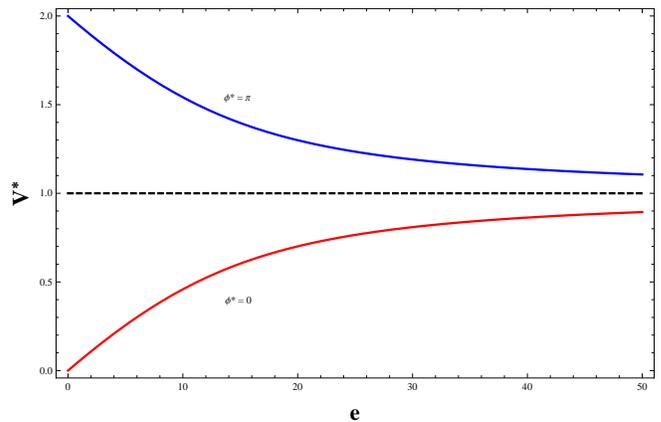}	
	\caption{Plot of $V^*$ vs $e$ for fixed $\omega$(shown for $\omega = 3$). The dotted line($V^* = 1$) separates the basin boundary for fixed points $A$(red curve for $\phi^* = 0$) and $B$(blue curve for $\phi^* = \pi$).}
	\label{Fig 6}
	\end{figure}

	For a given value of $\omega^2$, we show the variation of $V^*$ as a function of $e$ in Fig. \ref{Fig 6}.
	
	\vspace{.1in}
	
	The basin of attraction of $\phi^* = 0$ and $\phi^* = \pi$ are separated by the existence of the fixed point $(C)$. What we learn from Fig. \ref{Fig 6} is that the basin of attraction of $\phi^* = 0$ is confined to initial values of $V$ which are less than unity and the basin of attraction of $\phi^* = \pi$ is confined to initial values of $V$ which are greater than unity. Thus there are two non-overlapping segments of initial conditions which are driven to the two different fixed points.		
	
	\section{Acknowledgements}
	
	We would like to express our gratitude to The Department of Theoretical Physics, Indian Association for the Cultivation of Science, Jadavpur, Kolkata, India for providing support in conducting our research work.

	\end{document}